\documentclass[10pt,a4paper]{article}
\usepackage{graphicx}
\DeclareGraphicsExtensions{.jpg, .pdf, .png}

\date{}
\begin{document}

\date{}

\title{A New kinetic model of Globular clusters\\
{\normalsize }}
\author{Ll. Bel\thanks{e-mail:  wtpbedil@lg.ehu.es}}

\maketitle

\begin{abstract}

A new kinetic model of globular clusters based on a modified velocities distribution function is compared to the most often used King's model. A hypothetical contribution of dark matter is considered.
\end{abstract}

\section{Kinetic globular clusters models}

Globular clusters of stars are, roughly speaking, spherically symmetric aggregates of numerous stars maintained in a  volume of finite radius $a$ by their own gravitational field as they move around the galaxy to which they belong. Typical values are $N=10^4..10^6$ stars and $a=7..120$ {\bf pc}. Simple theoretical models, as the renowned King model \cite{King} that we review below, assume that i) all stars in the aggregate have the same mass, ii) that the mean gravitational field is static and iii) that the distribution of velocities $v^i$ (i=1,2,3) at each point of the cluster is isotropic. It follows from these three assumptions that it is possible to define for each point of the cluster a scape speed $v_e(r)$ where  $r$ means distance to the center of the cluster, such that beyond it the star would scape from the volume of radius $a$. This is also in particular the velocity that a star moving in the radial direction outwards should have to reach the frontier $a$ with zero velocity. This justifies the next assumption iv) that assumes that no stars with speeds greater than $v_e(r)$ remain in the cluster.

Both models of globular clusters considered here have as basic ingredient the fact of describing the gravitational mean field by a potential $V(r)$ solution of the Poisson equation:

\begin{equation}
\label{Poisson}
\frac{d^2V}{dr^2}+\frac{2}{r}\frac{dV}{dr}=4\pi G\mu
\end{equation}
where $\mu(r)$ must be chosen as a convenient description of the density of the ensemble of the stars of the cluster. This implies that at each point where there is a star, with velocity $v^i$ this star will move according to Newton's law:

\begin{equation}
\label{Newton}
\frac{dv^i}{dt}=-\delta^{ij}\frac{\partial V}{\partial x^j}
\end{equation}
and that:

\begin{equation}
\label{Energy}
E=\frac12 v^2+V(r)
\end{equation}
will be a constant of the motion.

Since the potential is defined only up to an arbitrary constant it is convenient to choose this constant so that $v_e(0)=-(-2V(0))^{1/2}$. It follows then from the assumptions i), ii) and iii) above the general relationship:

\begin{equation}
\label{Scape}
v_e(r)=(-2V(r))^{1/2}
\end{equation}
A nice consequence of it is that the boundary of the star cluster will coincide with the sphere $V(a)=0$

A kinetic globular cluster model is based on a choice of the  distribution function of the velocities of the stars of the cluster at each point of it, that because of the spherical symmetry  will be a function $f(r,v_i)$; and still more specifically, because of the assumption iii) above, it will be
a function $f(r,v)$ of two arguments only. The density of the cluster is then defined as:

\begin{equation}
\label{density}
\mu(r)=4\pi\mu_0 \int_0^{v_e(r)} f(r,v)v^2\,dv
\end{equation}

The initial choice that Chandrasekhar made was to choose the Maxwell-Boltzmann distribution function:

\begin{equation}
\label{Maxwell}
f(r,v)=ke^{-2j^2(v^2+2V(r))}
\end{equation}
$j$ being a constant with dimensions L${}^{-1}$T and $k$  a dimensionless normalization constant.
But keeping the range of $v$ being $v=0..\infty$, this led to models with infinite size and mass.

King models take care of this problem choosing the distribution function;

\begin{equation}
\label{King}
f(r,v)=ke^{-2j^2(V(r)-V(0))}(e^{-j^2v^2}-e^{2j^2V(r)})
\end{equation}
and restricting at each point $r$ the range of $v$ to be $v=0..v_e(r)$.
This leads to models with finite sizes and finite mass. But as we show below the same goal can be reached with the Maxwell-Boltzmann distribution function.

King's model deals with the problem of the scape velocity already at the level of the choice of the distribution function but abandons the Maxwell-Boltzman distribution function derived from fundamental classical statistics mechanics postulates. My model, based on (\ref{Maxwell}) delays dealing with the scape velocity condition until the definition of the density (\ref{density}) is used.

The constant $k$ being chosen for convenience so that:
\begin{equation}
\label{k}
4\pi\int_0^{v_e(0)}f(0,v)v^2\, dv=1
\end{equation}
the models to be discussed depend on three parameters: $j$, the central density $\mu_0$ and the central potential $V_0$, the derivative of the potential at the center being zero as a regularity condition. From these three parameters the first two are observable positive quantities with independent physical dimensions. Therefore we can assume that:

\begin{equation}
\label{units}
j=1, \quad \mu_0=1, \quad 4\pi G=1
\end{equation}
the last condition completing a convenient model dependent system of units.
Since $k$ given by the normalization condition (\ref{k}) is a function of $V_0$:

\begin{equation}
\label{kIII}
k=\frac{3}{\pi}\left(-6(-2V_0)^{1/2}e^{2V_0}+3\pi^{1/2}erf((-2V_0)^{1/2})-4(-2V_0)^{3/2}e^{2j^2V_0}\right)^{-1}
\end{equation}
for King's models; and:

\begin{equation}
\label{kI}
k=-\frac{e^{V_0}}{\pi}\left((-4j^2V_0)^{1/2}e^{V_0}-\pi^{1/2}erf(\frac12(-4j^2V_0)^{1/2})\right)^{-1}
\end{equation}
for the New ones, this leaves $V_0$ as the only one free parameter.

The Poisson equation (\ref{Poisson}) with the integral definition of either (\ref{King}) or (\ref{Maxwell}) expressions is a differential equation that can be dealt with conveniently writing it as the following system:

\begin{equation}
\label{System}
\frac{dV}{dr}=DV, \quad \frac{dDV}{dr}+\frac{2}{r}DV=\mu, \quad \frac{d\mu}{dr}=D\mu \\
\end{equation}
where:

\begin{equation}
\label{DmuK}
D\mu=-2\pi kDV(-2e^{2V_0}(-2V)^{1/2}+\pi^{1/2}e^{-2(V-V_0)}erf((-2V)^{1/2})
\end{equation}
for King models. And:

\begin{equation}
\label{DmuN}
D\mu=-\pi^{3/2} kDV e^{-V}erf((-V)^{1/2})
\end{equation}
for the New models.

I have integrated numerically the system of differential equations (\ref{System}) using Maple 16, with two different sets of six different values for the potential at the origin $V_0$. One set for King's models and the other for the New Models. Starting with $r=0$ I have recorded the values of the quantities $V(r)$, DV(r), and $\mu(r)$ until $r=b$ where:

\begin{equation}
\label{b}
\left(\frac{d^2V}{dr^2}\right)_{r=b}=\left(\mu-\frac{2}{r}DV\right)_{r=b}=0.
\end{equation}
$b$ is a distinguished value of $r$ where the force on any star of the cluster towards the center begins to decrease, and the left of $b$ defines the core of the cluster.
The integration proceeded then until both $\mu$ and $V$ reach the value zero at some point $a$, the radius of the cluster, also called the tidal radius. The solution to right of $b$ defines the halo of the cluster to be matched with a vacuum solution at $r=a$.

The results of these integrations are shown in the table below for King's models

\vspace{1cm}
\begin{tabular}{|l|l|l|l|}
\hline
\\[-2ex]
$\quad V_0$ & $\quad b $ & $\quad a$ & $\quad c $
\\[.5ex]
\hline
-1 & \ 1.4915 & \ 6.753 & \ 0.66\\
-2 & \ 1.9085 & \ 14.515 & \ 0.88\\
-3 & \ 2.068 & \ 36.8 & \ 1.25\\
-4 & \ 2.115& \ 127 & \ 1.78\\
-5 & \ 2.124 & \ 360 & \ 2.22\\
-6 & \ 2.127 & \ 750 & \ 2.55\\
\hline
\end{tabular}
\begin{tabular}{|l|}
\hline
\\[-2ex]
$\quad b_{1/3}$
\\[.5ex]
\hline
1.5735 \\
1.9787 \\
2.120 \\
2.1585 \\
2.1674 \\
2.1690 \\
\hline
\end{tabular}
(King)
\vspace{1cm}

and in the table below for the New models:

\vspace{1cm}
\begin{tabular}{|l|l|l|l|}
\hline
\\[-2ex]
$\quad V_0$ & $\quad b $ & $\quad a$ & $\quad c $
\\[.5ex]
\hline
-1 & \ 1.7905 & \ 4.05009 & \ 0.35\\
-3 & \ 2.702 & \ 9.6739 & \ 0.55\\
-5 & \ 2.948 & \ 21.285 & \ 0.86\\
-7 & \ 2.998 & \ 57.26 & \ 1.28\\
-9 & \ 3.006 & \ 180.7 & \ 1.78\\
-11 & \ 3.007 & \ 501 & \ 2.22\\
-13 & \ 3.008 & \ 1220 & \ 2.61\\
\hline
\end{tabular}
\begin{tabular}{|l|}
\hline
\\[-2ex]
$\quad b_{1/3}$
\\[.5ex]
\hline
1.9197 \\
2.809\\
3.02 \\
3.06 \\
3.0665\\
3.068 \\
3.068\\
\hline
\end{tabular}
(New)
\vspace{1cm}

The fourth column of the tables above lists the values of the quantity:

\begin{equation}
\label{c}
c=\log_{10}\left(\frac{a}{b}\right)
\end{equation}
and is the analogous of the quantity $\log_{10}(a/r_c)$ that King calls the concentration of the cluster. This parameter is important because it is dimensionless and therefore it is not affected by the object dependent system of units defined by (\ref{units}).
To compare King models to the New ones in the range of $c$ considered in the preceding tables I it is useful to realize that in both cases the dependence of $c$ on $V_0$ can be fitted  approximately by linear equations:

\begin{equation}
\label{}
c=0.20-0.38\, V_0 \ \hbox{(King)}  \quad c=-0.15-0.21\, V_0 \ \hbox{(New)}
\end{equation}
This allows to choose , by interpolation, pairs of potentials that will lead to the same concentration parameter $c$.

The fifth column lists the values of $r$ where the value of $\mu$ is one third of $\mu_0$.

King defined the core radius\footnote{Observationally  the core radius is the distance at which the apparent surface luminosity has dropped by half} of the cluster empirically as the value of $r$ such that:

\begin{equation}
\label{rc}
8\pi Gj^2r_c^2\mu_0=9
\end{equation}
so that with the units that I am using (\ref{units}) we have $r_c=3/(2)^{1/2}\simeq 2.12$
that approximates but does not coincide with the parameter $b$ which has a clear physical meaning indicating where the force that the gravitational attracting force towards the center of the cluster starts to decrease.

\vspace{1cm}
\begin{tabular}{|l|l|l|l|l|}
\hline
\\[-2ex]
$\quad c $ & $\quad b_K $ & $\quad b_N $ & $\quad a_K $ & $\quad a_N$
\\[.5ex]
\hline
0.7(2+1) & \ 1.7653 & \ 2.891 & \ 8.91 & 15.409\\
0.9(7+1) & \ 1.974 & \ 2.973 & \ 18.44 & 28.155.\\
1.7(5+1) & \ 2.113 & \ 3.006 & \ 120. & 172.8.\\
2.0(8-3) & \ 2.165& \ 3.007 & \ 250 & 337.\\
2.5(1-1) & \ 2.169 & \ 3.008 & \ 680 & 960.\\

\hline
\end{tabular}
\vspace{1cm}

The subindex $K$ refers to a King model and $N$ to a New one. The $+$ and $-$ in the last column is a correction to get the corresponding values for the New models. A general feature appears from looking at this table, namely  that with equal concentration $c$ the New models are bigger than King models.

Let us assume that we know the values of the density $\mu^*_0$ at the origin of a real cluster and the value of $j^*$ using a cluster independent system of units; for instance the MKS system of units. Then the units of the system for which the conditions (\ref{units}) will hold are:

\begin{equation}
\label{Units transform}
m^*=\frac{1}{2j^*} (\pi G^*\mu^*_0)^{-1/2}, \quad kg^*=\frac{mu^*}{8{j^*}^3} (\pi G^*\mu^*_0)^{-3/2}, \quad
s^*=\frac{1}{2} (\pi G^*\mu^*_0)^{-1/2},
\end{equation}
And therefore, in particular,  the value of the parameters with dimension L in the preceding tables,i.e: $a$, $b$ and $b_{1/3}$ will be multiplied by $m^*$ in the cluster independent system of units. This is a step that it is necessary to compare theoretical models with observations of existent globular clusters. But To this end a problem remains, namely to establish a correspondence between the theoretically defined parameters $b$ or $b_{1/3}$ with optically defined characteristics of the globular clusters, and this is beyond the scope of this paper.

The following graphics below complete the comparison of the two models for the case $c=1.78$ that is present in both tables above.


\section{Dark matter}

In a preceding paper, \cite{Bel}, I suggested that a very simple, though unacceptable {\it a priori} to some colleagues,  modification of Newton's action at a distance law, could solve the problem of dark matter, or at least mimic what dark matter is supposed to be. The idea is to add to the $1/r^2$ term of the law  a $1/r$ one so that the force between two point particles of masses $m_1$ and $m_2$ at a distance $r$ becomes:

\begin{equation}
\label{Newton}
F+F^\prime, \  \hbox{with} \ F=-\frac{Gm_1m_2}{r^2}, \ F^\prime=-\frac{G^\prime m_1m_2}{r}
\end{equation}
where $G^\prime$ would be a new universal constant with a value small enough so that the new law would not contradict any observationally known result.

To discuss the range of possible values of $G^\prime$ it is more intuitive to deal with the following  constant:

\begin{equation}
\label{beta}
\beta=\frac{4\pi G}{G^\prime}
\end{equation}
that has physical dimension L and is the inverse of $G^\prime$ when using the condition $4\pi G=1$. If $r=\beta/4\pi$ then $F=F^\prime$, if $r<\beta/4\pi$ then $F>F^\prime$ and $F<F^\prime$ otherwise.

It is also useful to realize that the above modification of Newton´s law is equivalent to assuming that $G$ instead of being a constant is in fact a function of $r$:

\begin{equation}
\label{G(r)}
G(r)=G_0(1+r\beta^{-1})
\end{equation}
In a Cavendish-like experiment $r$ is of the order of 10 cm and therefore if $\beta$ is greater than 100 km  then:

\begin{equation}
\label{G0}
\frac{G-G_0}{G_0}< 10^{-7}
\end{equation}
that is compatible with the uncertainty of the present measured value of $G=6.67384(80)\,\times 10^{-11}$ (SI units).

We deal with gravity organized systems at increasing scales. From planets and the solar system I take for granted, as a guess, that
$\beta$ greater than 1000 a.u. would be compatible with any known facts. And so the next scale that I consider is that of  globular clusters.

From what I wrote in \cite{Bel} it follows that a spherical distribution of mass with density $\mu(r)$ contained in a volume of radius $a$ will generate, in the modified Newton's theory, a gravitational field equivalent to that generated by the classical one if the density were $\rho(r)=\mu(r)+\bar\mu(r)$, with:

\begin{equation}
\label{Modified1}
\bar\mu(r)=2\pi\beta^{-1}\int_0^a du\int_0^\pi d\theta\frac{\mu(u)u^2\sin \theta}{r^2+u^2-2ru\cos \theta },
\end{equation}
that after the $\theta$ integration becomes:

\begin{equation}
\label{Modified2}
\bar\mu(r)=\pi\beta^{-1}\int_0^a du \frac1r\mu(u)u\ln\frac{(r+u)^2}{(r-u)^2};
\end{equation}
To develop a theory of globular clusters taking in consideration the effective dark matter density
would require to use the composite density $\rho(r)$ instead of $\mu$, (\ref{System}) becoming an integral-differential system of equations. This looks daunting and totally unnecessary at this stage of the theory.

What is possible and eventually interesting at this point is to choose an appropriate value for $\beta$ and consider $\bar\mu(r)$ as a small fraction of $\mu(r)$ so that (\ref{System}) can be considered a first approximation to a more complex problem, that  give  us a hint on whether dark matter may be present in globular clusters. I offer the figures below to help the reader make his/her mind. Notice that for graphical readability I have chosen $\beta$ equal $3a$ for the King model and $\beta$ equal $a$ for the New model. At the theoretical level of this paper $\beta$ is a free parameter that only observations could fix.

\section*{Aknowledgements}

Alfred's Latex expertise is gratefully acknowledged. Alberto and Juan Mari both helped me to improve the manuscript.

\begin{figure}[htbp]
\begin{center}
\begin{minipage}[b]{15cm}
\includegraphics[width=7cm]{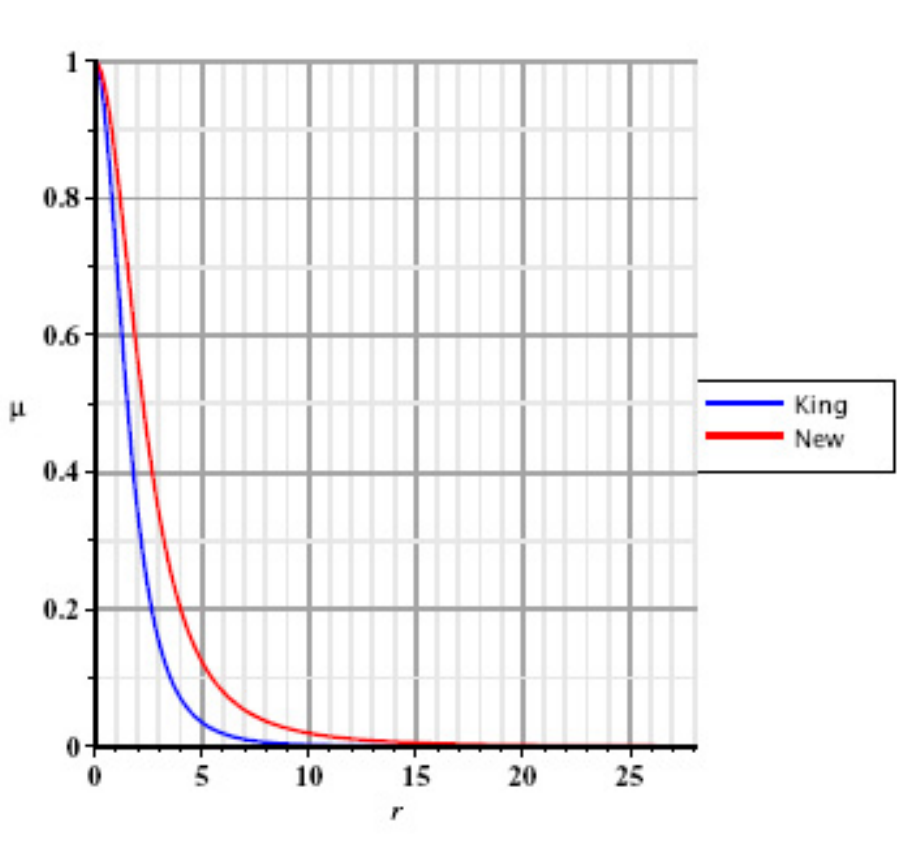}
\includegraphics[width=7cm]{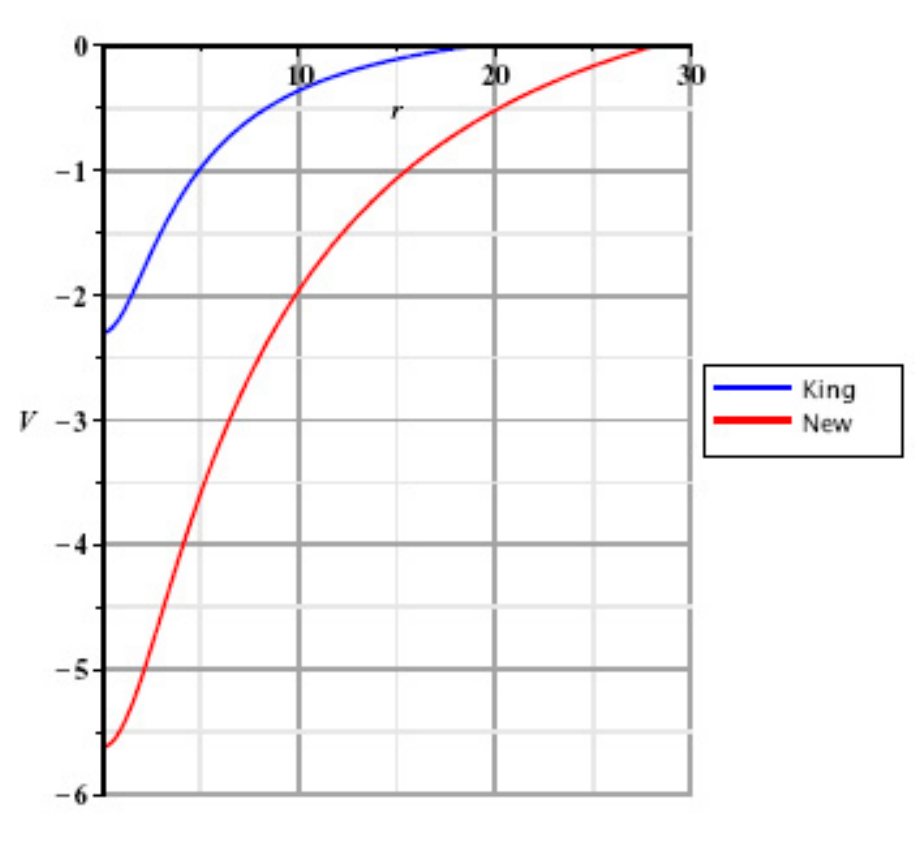}
\end{minipage}
\end{center}
\end{figure}

\begin{figure}[htbp]
\begin{center}
\begin{minipage}[b]{15cm}
\includegraphics[width=7cm]{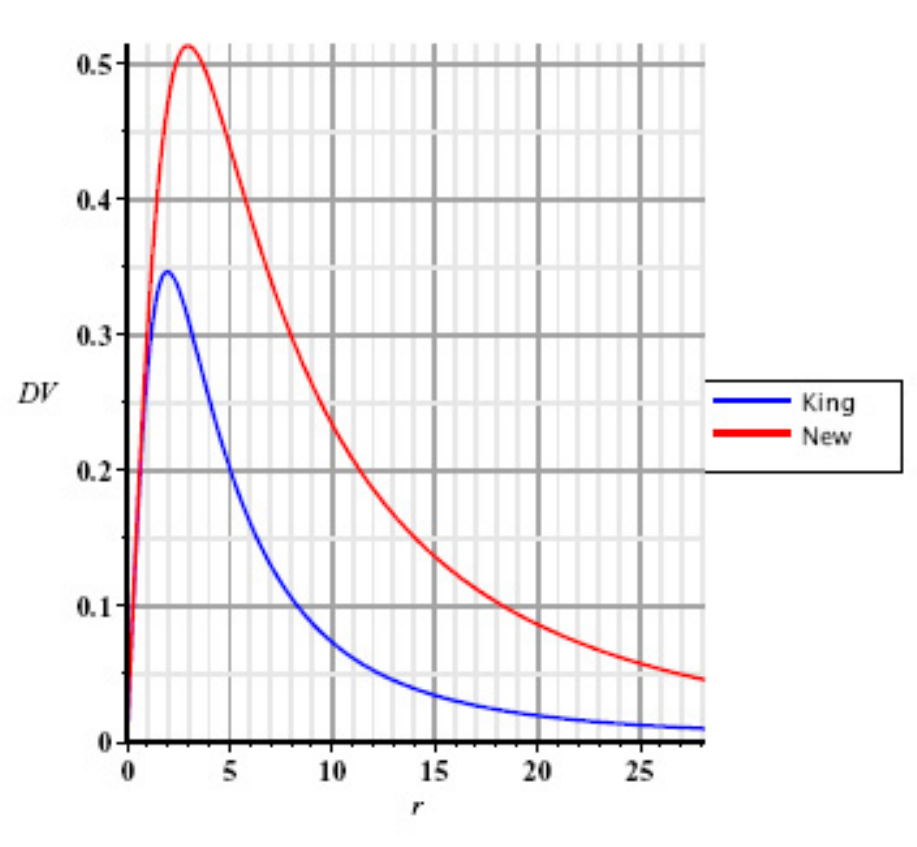}
\includegraphics[width=7cm]{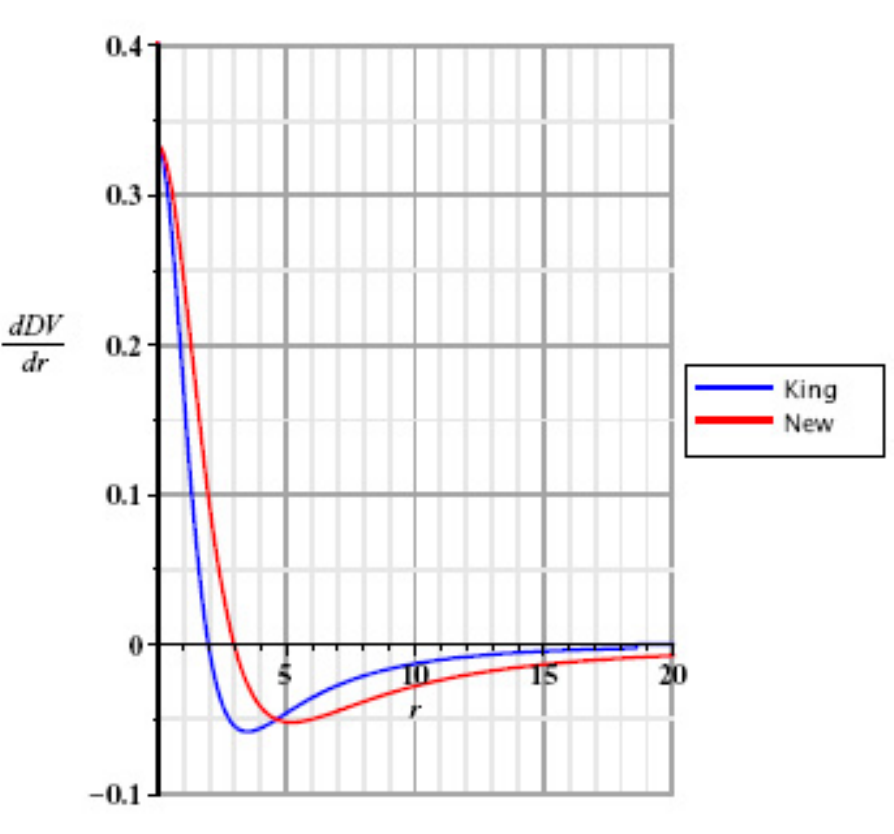}
\end{minipage}
\end{center}
\end{figure}

\begin{figure}[htbp]
\begin{center}
\begin{minipage}[b]{15cm}
\includegraphics[width=7cm]{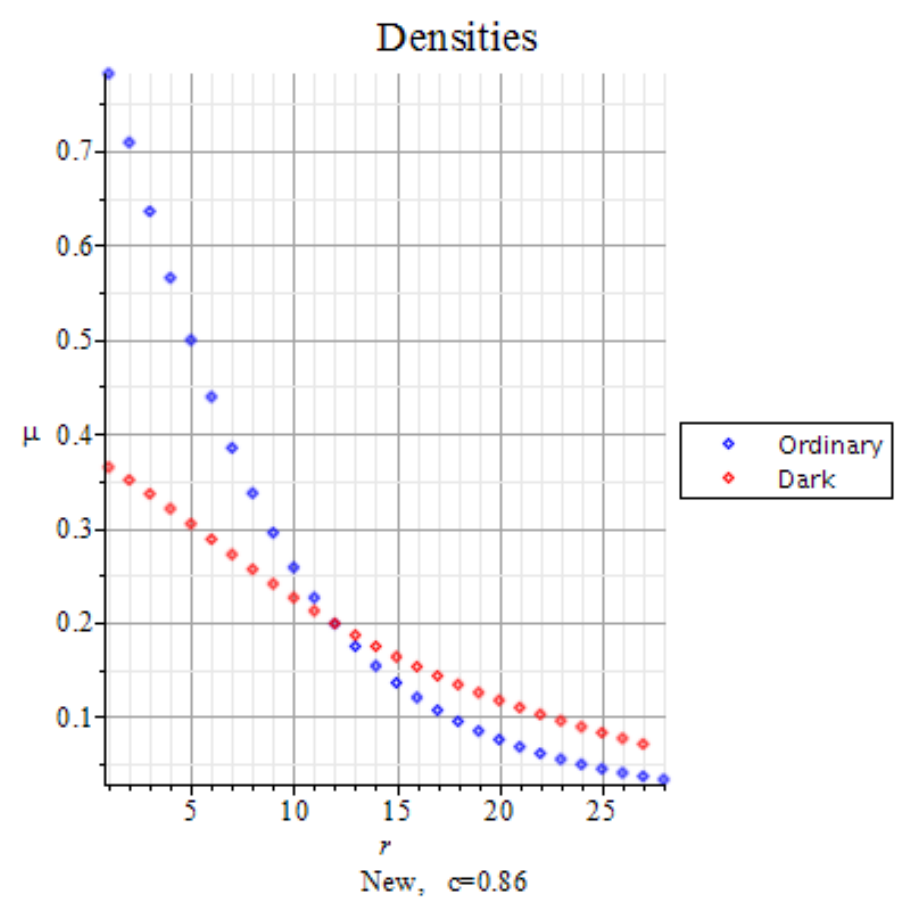}
\includegraphics[width=7cm]{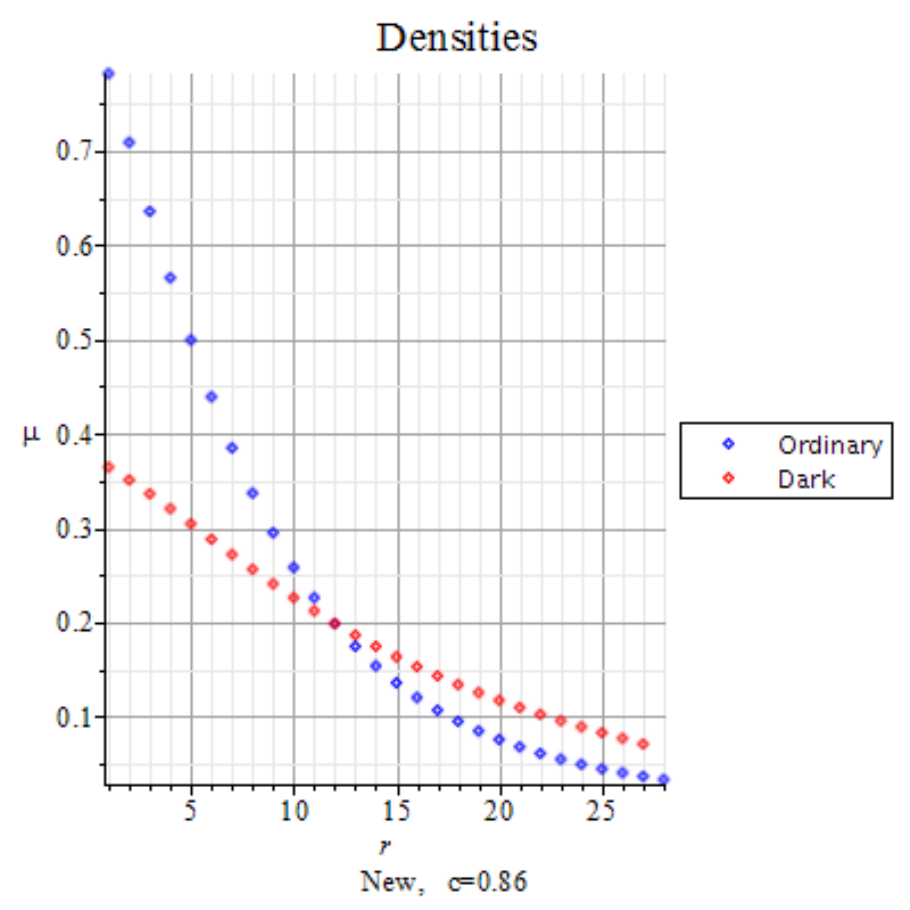}
\end{minipage}
\end{center}
\end{figure}


\begin{thebibliography}{9}
\bibitem{King} I.\ R.\ King, Astro. Journ., {\bf 71}, 1, p.64 (1966)
\bibitem{Bel} Ll. Bel,  arXiv:1308.0249v2 [physics.gen-ph]

\end{thebibliography}
\end{document}